\RequirePackage{fix-cm}
\documentclass[]{svjour3}                     
\smartqed  
\usepackage{graphicx}
\usepackage{amsmath}
\usepackage{array,multirow,setspace}
\usepackage{pgfplots}
\usepackage[utf8]{inputenc}
\usepackage{tipa}
\usepackage[autostyle]{csquotes}
\usepackage[T1]{fontenc}
\pgfplotsset{compat=1.7}
\usepackage{mathtools}
\usepackage{caption}
\usepackage{comment}
\usepackage{makecell}
\usepackage{arydshln}

\DeclareUnicodeCharacter{2212}{-}

\begin{document}

\title{Prosody-TTS: An end-to-end speech synthesis system with prosody control}  

\author{Giridhar Pamisetty         \and
        K. Sri Rama Murty}
\institute{Giridhar Pamisetty \at
              Speech Information Processing lab \\
              Department of Electrical Engineering\\
              Indian Institute of Technology Hyderabad, India\\
              \email{ee17resch11001@iith.ac.in}
           \and
           K. Sri Rama Murty \at
              Speech Information Processing lab \\
              Department of Electrical Engineering\\
              Indian Institute of Technology Hyderabad, India\\
              \email{ksrm@ee.iith.ac.in} \\
}

\date{Received: date / Accepted: date}

\maketitle

\begin{abstract}

End-to-end text-to-speech synthesis systems achieved immense success in recent times, with improved naturalness and intelligibility.  
However, the end-to-end models, which primarily depend on the attention-based alignment, do not offer an explicit provision to modify/incorporate the desired prosody while synthesizing the signal.
Moreover, the state-of-the-art end-to-end systems use autoregressive models for synthesis, making the prediction sequential. Hence, the inference time and the computational complexity are quite high.
This paper proposes Prosody-TTS, an end-to-end speech synthesis model that combines the advantages of statistical parametric models and end-to-end neural network models. It also has a provision to modify or incorporate the desired prosody by controlling the fundamental frequency ($f_0$) and the phone duration.
Generating speech samples with appropriate prosody and rhythm helps in improving the naturalness of the synthesized speech.
We explicitly model the duration of the phoneme and the $f_0$ to have control over them during the synthesis.
The model is trained in an end-to-end fashion to directly generate the speech waveform from the input text, which in turn depends on the auxiliary subtasks of predicting the phoneme duration, $f_0$, and mel spectrogram.
Experiments on the Telugu language data of the IndicTTS database show that the proposed Prosody-TTS model achieves state-of-the-art performance with a mean opinion score of 4.08, with a very low inference time.

\end{abstract}

\section{Introduction}
\label{intro}
Text-to-Speech Synthesis (TTS) is the process of generating the natural-sounding expressive speech from the given text. Although several techniques were proposed for speech synthesis, generating natural-sounding speech signals is still a challenging task. The allophonic variations and coarticulation make it challenging to generate the natural-sounding speech from the text \cite{Allophones}. Major applications of the TTS system include human-machine interactions, speech-to-speech translation (automatic dubbing), screen readers, etc., where the speech samples have to be synthesized with the desired prosody and emotion. Duration of the phones, fundamental frequency $(f_0)$, and the energy of the speech signal are considered as the crucial parameters responsible for the prosody and the emotion in the speech signal \cite{emo_para}. So, having explicit control over these crucial parameters helps in synthesizing the speech signal with desired prosody and emotion. For example, in the task of automatic dubbing \cite{machine_dubbing}, the source language speech duration should be incorporated in the target language speech to match with the events in the source video. So, we need to have control over the predicted durations to alter them during the synthesis. Most state-of-the-art end-to-end models do not have the provision to control the duration of the phones, fundamental frequency, and energy. In this work, we propose an end-to-end speech synthesis system with a provision to control the duration and the $f_0$, which would better suit for real-time practical applications like expressive conversational speech synthesis, automatic dubbing (subtitle synthesis), emotion conversion, etc.

Before the advent of deep neural networks (DNNs), unit selection \cite{Unit_Sel} and statistical parametric speech synthesis (SPSS) \cite{SPSS} were the most successful speech synthesis frameworks. Unit selection based speech synthesis is a data-driven approach that generates the speech signal by concatenating the prestored temporal waveforms of the subword units like phonemes and syllables, based on the target and the concatenation cost. The database should be large enough to cover all the possible allophonic variations and coarticulation contexts of the sound units. The performance of the unit selection system degrades when there is a domain mismatch or when the context of the inference text doesn't occur in the training data. This system doesn't offer control over the prosodic parameters as it concatenates the prestored temporal waveforms during the synthesis.

SPSS system has several hand-engineered modules such as a text analyzer for linguistic feature extraction, a duration model to predict the duration of each phoneme, an acoustic model to estimate the acoustic parameters from the linguistic features, and finally, a vocoder to synthesize the speech signal from the predicted acoustic parameters. Instead of storing the temporal waveforms of the phonemes/syllables, the SPSS system models the speech parameters using stochastic generative models like the hidden Markov model (HMM). In the HMM-based SPSS \cite{HTS}, the context-dependent HMMs are used to model the probability density function (pdf) of acoustic parameters conditioned on linguistic contexts such as parts of speech, tone, lexical stress, and pitch. The HMM states are modeled as Gaussian density functions parameterized by the mean vectors and covariance matrices. The duration of the phonemes is modeled by the transition probabilities of the HMM. During synthesis, the acoustic (spectral and excitation) parameters are generated by sampling from the HMM states \cite{para_gen}. Since the SPSS system models $f_0$ as one of the acoustic parameters, it can be controlled/modified during the synthesis. The SPSS system can generate the speech from the unseen context in the inference text as it models the statistical properties of the parameters.  But, the generated excitation and spectral parameters from the HMMs are often over-smoothed, making the synthesized speech sound muffled. The major drawback of the SPSS system is the quality of synthesized speech, which is affected due to over smoothing, the accuracy of the acoustic model and the vocoder \cite{SPSS}.

In order to overcome the issues involved with HMM-based acoustic modeling, neural network models were employed to capture the non-linear relationship between the linguistic and the acoustic features. Neural parametric speech synthesis systems like Merlin \cite{Merlin} contains similar modules as the SPSS system, with the duration and acoustic models replaced with simple feed-forward neural networks. The neural network models are trained to estimate the acoustic parameters from the context-dependent linguistic features extracted from the text. Usually, pentaphone context, with two phones on either side of the current phone, is used to predict the duration of the phoneme \cite{Merlin}. The predicted durations are used to upsample the linguistic features to the acoustic frame rate and fed to the acoustic model. The speech parameters predicted from the acoustic model are given to the vocoder to synthesize the speech signal.

The Merlin system can be trained with relatively lesser data (3 to 4 hours) as it uses explicit alignments obtained from the duration model. The overall performance of the Merlin system depends on the accuracy of all individual modules, as they are trained separately. The intelligibility of the synthesized samples has improved when compared to the SPSS system. Although the neural parametric speech synthesis systems have the provision to change or incorporate the desired prosody by modifying the $f_0$ and phoneme duration, the synthesized samples are not natural because of the lack of coordination between individual modules-Text analyzer, duration model, acoustic model, and vocoder. The signal processing based vocoders rely on several assumptions in modeling human speech production mechanism as the source-filter model, which causes the degradation in the naturalness of the synthesized samples. Parametric speech synthesis systems require a language-dependent text analyzer to extract the linguistic features. But, most languages do not have better text analyzers to extract reliable linguistic features from the text.

With the development of DNNs in the recent times, several end-to-end speech synthesis models such as Tacotron \cite{Tacotron}, Transformer TTS \cite{TrTTS}, Deep Voice \cite{DeepVoice}, Char2wav \cite{Char2Wav}, Clarinet \cite{ClariNet} etc., are developed. The end-to-end models do not require explicit extraction of the context-dependent linguistic features from the text. Instead, they learn them from the text data using a text encoder module. They convert the input text into character embeddings and use a text encoder module to encode them into a representation, which captures the required context information. They rely on the attention mechanism to learn the alignment between linguistic and acoustic features, as they operate at different rates. An autoregressive decoder estimates the linear spectrogram or the mel spectrogram frame by frame from the encoded representations, making the prediction sequential \cite{Tacotron}. The speech signal is generated from the predicted spectrograms using the signal processing based vocoders like Griffin-Lim \cite{GriffinLim}, or neural vocoders like WaveNet \cite{Wavenet}, WaveRNN \cite{WaveRNN}, etc.

The synthesized speech samples from the end-to-end models are highly intelligible with improved naturalness, but the inference time is very high because of the autoregressive architecture. We also observed the problem of missing words while synthesizing longer utterances and repetition of words while synthesizing shorter utterances, which could be attributed to the maximum decoder steps hyper-parameter in the attention-based alignment\footnote{Samples are available at: https://siplabiith.github.io/prosody-tts.html}. Although the end-to-end models do not require the language-dependent text analyzer, they need a huge amount of training data, about 25 to 40 hours, to extract context-dependent information and align the linguistic and acoustic features. Furthermore, they do not offer a provision to control the duration of the phoneme and the $f_0$.

We propose Prosody-TTS, a hybrid model that combines the advantages of SPSS and end-to-end neural synthesizers to achieve naturalness with lesser data and lower inference time. Further, having control over the phoneme duration and $f_0$ makes the TTS systems useful for various applications like emotion conversion, automatic dubbing, expressive speech synthesis, etc. So, instead of using the attention-based alignment in the proposed model, we explicitly model the duration of the phonemes to have control over it. The $f_0$ of each frame is also modeled explicitly and used for conditioning the mel spectrogram prediction and to create the excitation input signal for the neural vocoder \cite{waffler}.

The Prosody-TTS system generates the speech signal from the input text sequence, represented as fixed non-learnable character embeddings. The model is trained end-to-end using supervised auxiliary learning since it improves the ability of the model to generalize on the unseen data \cite{Aux_task1,go_deep}. The individual modules (duration estimator, $f_0$ estimator, acoustic decoder, and neural vocoder) in the proposed architecture are considered as the subtasks, whereas the primary task is to predict the speech samples from the character embeddings of the given text. The major limitation of supervised auxiliary learning is obtaining the auxiliary target data for each subtask. The proposed model overcomes this limitation inherently since each subtask has the corresponding target data (duration, $f_0$, mel spectrogram, etc.). Each subtask is intended to perform a certain function, allowing us to improve a specific module easily. The salient features of the Prosody-TTS model are:
\begin{itemize}
    \item Duration and $f_0$ are explicitly modeled, as they can be controlled to modify the prosody or incorporate the desired prosody.
    \item Requires a significantly less amount of training data, which helps in building the TTS systems for low resource languages.
    \item An end-to-end model with a very low real-time factor of 0.13
\end{itemize}

The rest of the paper is organized as follows: The architecture of the Prosody-TTS system, along with the prosody incorporation, is explained in section \ref{sec:Nw_Arch}. Section \ref{sec:Exper_setup} describes the speech synthesis experiments and compares the Mean opinion scores (MOS) of synthesized speech samples with different state-of-the-art models. Section \ref{pros_cont} describes the proposed prosody control method and compares the MOS of prosody modified and prosody incorporated samples with the other systems. Section \ref{sec:Conclusion} concludes the paper with few insights about the future work.

\section{Architecture of Prosody-TTS}
\label{sec:Nw_Arch}
\begin{figure*}[t]
    \centering 
    \includegraphics[width=\textwidth]{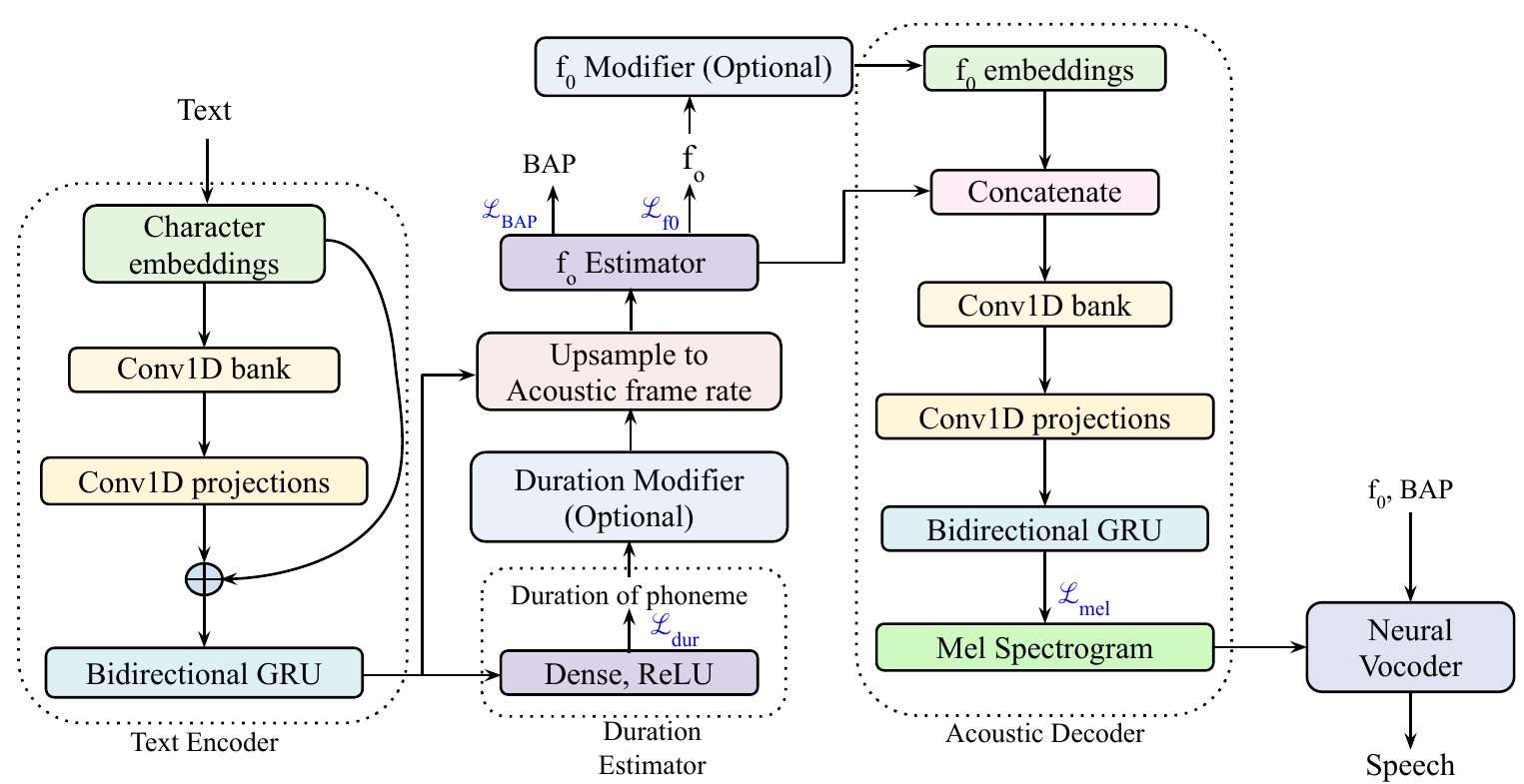}
    \captionsetup{justification=centering}
    \caption{Architecture of the Prosody-TTS system}
    \label{arch}
\end{figure*}

The Prosody-TTS system uses a hybrid approach by utilizing the advantages from the parametric synthesis models and the end-to-end models. The parametric speech synthesis models require a lesser amount of training data as the phoneme durations are explicitly modeled. Although the inference time and computational complexity are very low, the lack of coordination between the individual modules and the accuracy of the vocoder makes the synthesized samples sound unnatural. The end-to-end model uses attention-based alignment, which requires a larger amount of training data. But, the model is trained in an end-to-end fashion, which improves the naturalness of the synthesized samples. So, advantages of the parametric and end-to-end models are exploited in developing the architecture of the Prosody-TTS system, which is shown in the figure \ref{arch}.

Prosody-TTS system consists of a text encoder module, duration estimator, $f_0$ estimator, acoustic decoder module, and a neural vocoder. The text encoder module encodes the input text sequence into hidden representations that capture the linguistic context information present in the text. The duration estimator uses these context-dependent encoded representations to estimate the duration of each phoneme in the given text. The text encoder module operates at the linguistic phoneme rate, while the acoustic decoder module operates at the acoustic frame rate. So, we upsample the bottleneck representations of the text encoder to the acoustic frame rate using the predicted durations and feed to the acoustic decoder. Upsampling is done by replicating encoded representations of each phoneme based on the predicted duration of the corresponding phoneme. The $f_0$ estimator takes the upsampled representations, predicts the $f_0$ and band aperiodicity (BAP) of each frame. $f_0$ estimator is followed by an acoustic decoder module, which generates the mel spectrogram of the speech signal by conditioning on the $f_0$. An excitation signal is created from the $f_0$ and BAP, which is fed to the neural vocoder along with the predicted mel spectrogram to generate the corresponding speech signal. The Prosody-TTS model uses a non-autoregressive neural vocoder \cite{waffler}. The model is trained in an end-to-end fashion using supervised auxiliary learning, with each module of the Prosody-TTS architecture acting as a subtask. The individual modules are briefly explained in the following subsections.

\subsection{Text-Encoder module}
The input text is converted to the international phonetic alphabets (IPA), which is a phonetic transcription of text and a standardized representation of speech sounds \cite{IPA}. Each character in the encoded IPA text is referred to as a phoneme. A 128-dimensional fixed and non-learnable random vector is used to represent each phoneme of the input text. In contrast to the fixed pentaphone context information in parametric methods, we use a text encoder module consisting of CNNs to model the coarticulations and allophonic variations of the speech as the CNNs capture the context information.

The phoneme representations are fed to the bank of one-dimensional convolutional filters with ReLU activations to capture the context information through the receptive field. In the Prosody-TTS, we use eight one-dimensional convolutional filters in the bank, where $k^{th}$ filter has the kernel size of k. Thus, the convolution bank captures the context information up to 8 phonemes as the maximum receptive field in the bank is 8. The output of each convolutional filter in the bank is stacked horizontally and projected into a fixed (lower) dimensional representation using two convolutional projection layers with a kernel size of 3. This results in the effective receptive field of 12, indicating that we provide a maximum of 12 phoneme context information. The input phoneme representations are added with the fixed dimensional representations through the residual connection to provide the lost phoneme identity and fed to the bidirectional gated recurrent unit (GRU).

\begin{figure}[t]
    \captionsetup{justification=centering}
    \centering
        \input{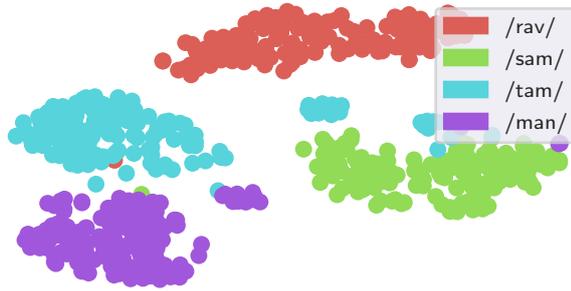}
    \caption{t-SNE plot of the encoded representations of phoneme /a/ \\ occurred in different contexts.}
    \label{tsne}
\end{figure}

Although the random embeddings are used for representing the phoneme, the text encoder module incorporates the context-dependent information in the encoded representations, which is required for modeling the durations and allophonic variations of the speech. For example, consider phoneme /a/, which is initially represented with a context-independent random vector, and they correspond to a point. But, the encoded representations from the text encoder module contain the phoneme context information. The t-SNE plot in the figure \ref{tsne} shows the encoded representations of /a/ in four different contexts (/sam/, /rav/, /tam/ and /man/). We can observe the similar context representations are grouped, indicating that the context information is captured. There is a slight overlap in the encoded representation of /a/ in the context of /sam/ and /tam/ as the right context phoneme is similar in both cases. We should also note that no loss function is involved in training the text encoder module, and it learns along with the other modules as we employ end-to-end training. Thus, we can eliminate the need for the text analyzer to extract linguistic features, which requires language-specific knowledge.

\subsubsection{Importance of contextual information in duration modeling}
\label{sec:Need of context info}

\begin{figure}[t]
    \captionsetup{justification=centering}
    \resizebox{\textwidth}{!}{%
    \centering
        \input{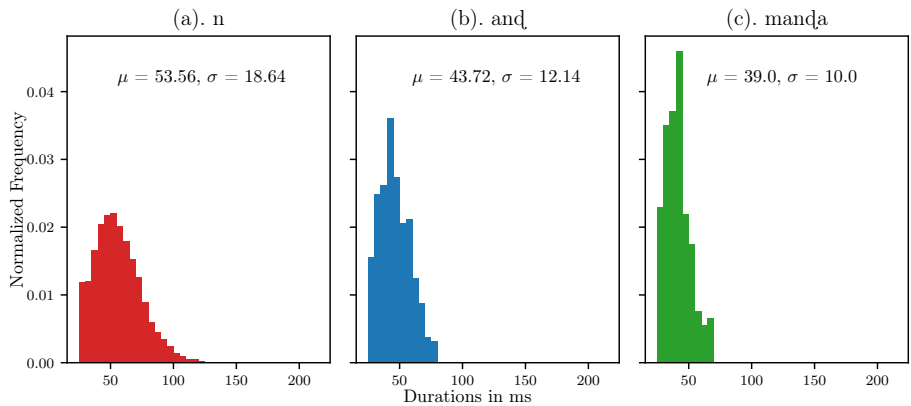}
    }
    \caption{Duration distribution of phoneme /n/ (a): without any context, (b): in the context of /a/ and /\textrtaild/, (c): in the context of /ma/ and /\textrtaild a/.}
    \label{distribution}
\end{figure}

Speech signal exhibits a high degree of temporal dependencies, and the prosody of the speech signal depends on the context in which the corresponding phonemes occur. Duration and intonation are the two crucial prosodic features responsible for generating natural sounding speech \cite{SLP}, which in turn depends on the contextual information. Thus, contextual information affects the prediction of phoneme durations and the naturalness in continuous speech. Figure \ref{distribution}(a) shows the distribution of durations (in ms) of the context-independent phoneme /n/. We can observe that the variance of the distribution is higher, with the minimum and maximum duration being 25 ms and 215 ms, respectively. Since the durations of the context-independent phoneme are distributed in a wider range, it is difficult to model and predict them accurately.

Figure \ref{distribution}(b) shows the distribution of durations of phoneme /n/ in the context of /a/ and /\textrtaild/, which are the left and right phonemes of the current phoneme /n/, respectively. We can observe that the durations are distributed in the smaller range, where the minimum and maximum durations are 25 ms and 90 ms, respectively. So, the addition of left and right context information makes the distribution sharper and reduces uncertainty.

Figure \ref{distribution}(c) shows the distribution of durations of phoneme /n/ in the context of /ma/ and /\textrtaild a/, which are the two left and the two right phonemes of the current phoneme /n/, respectively. There is a further decrease in the variance of the distribution, with the minimum and maximum values of 25 ms and 70 ms, respectively. Thus, contextual information reduces the variance of the distribution, which is essential for training the accurate duration model. In this paper, we use 12 phoneme context information captured in the text encoder for duration estimation. So, the uncertainty in predicting the phoneme duration reduces to a great extent. We didn't consider further higher context information, as the number of training examples is minimal for such higher contexts, and the model gets over-fitted to the training data. The proposed model is also evaluated with various phoneme contexts, and the results are compared in the experimental section.

\subsection{Duration estimator}
\begin{figure}[t]
    \resizebox{\textwidth}{!}{%
    \centering
        \input{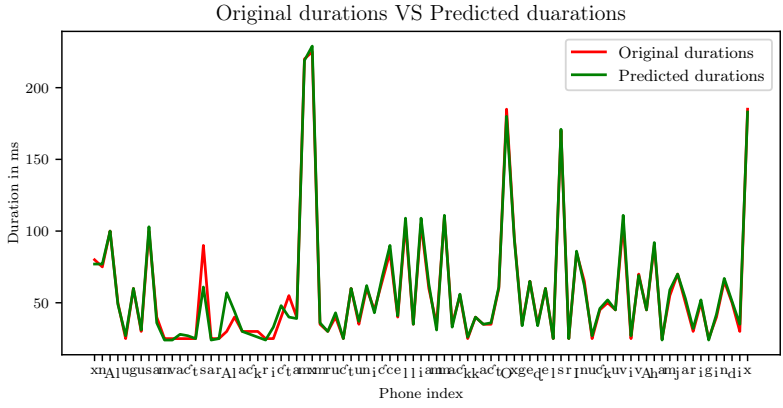}%
    }
    \caption{Comparison of the original and predicted durations of an utterance}
    \label{comp_durs}
\end{figure}

Let X = \{$x_1, x_2, x_3, \dots, x_N$\} be the input phoneme sequence of an utterance, E = \{$e_1, e_2, e_3, \dots, e_N$\} be the embeddings obtained from the text encoder module, where N is the total number of phonemes in the given utterance. Let T = \{$t_1, t_2, t_3, \dots, t_N$\} be the duration sequence of the phonemes, where $t_n$ is the duration of the phoneme $x_n$ in milliseconds (ms). The task of duration estimator is to predict the duration sequence $t_1, t_2, \dots, t_N$ from the embeddings $e_1, e_2, \dots, e_N$. Since the context information required for duration modeling is already captured in the text encoder, we use a simple affine transformation to estimate the durations. So, the duration estimator contains only a single dense layer with a ReLU activation function. The optimum values $(t^*)$ are found by reducing the mean absolute percentage error (MAPE) between the original and predicted durations. Arnaud et al. \cite{mape} showed the existence of an optimal model regarding MAPE and also proved the empirical risk minimization. This error particularly helps when the target variable is positive and quite far from zero, as it calculates the percentage of error. The stop consonants have shorter durations (around 50ms) when compared to the vowels (150-250ms) as they are impulsive in nature. So, vowels can have an error of around 30ms, but stop consonants can't afford such large errors. Thus, the error is made in proportion to their duration by using the percentage of error. The loss associated with the duration estimator is given by:
\begin{equation}
    \mathcal{L}_{dur} =  \frac{1}{N} \sum_{n=1}^N  \frac{|t_n - \hat{t}_n|}{t_n}
\end{equation}
where $\hat{t}$ are the predicted durations obtained from the duration estimator. Figure \ref{comp_durs} shows the comparison between the original and predicted durations for the utterance "nAlugu sa$\theta$va\texthtc tsarAla \texthtc kri\texthtc tam, mru\texthtc tuni \texthtc celli amma\texthtc kka\texthtc tO, gedel \textesh rInu\texthtc ku, vivAham jarigi$\theta$di." We can observe that the durations are much closer, which also demonstrates the efficiency of the text encoder.

\subsection{$f_0$ estimator}
\label{sec:f0_est_module}
\begin{figure}[t]
    \begin{center}
    \resizebox{\textwidth}{!}{%
        \input{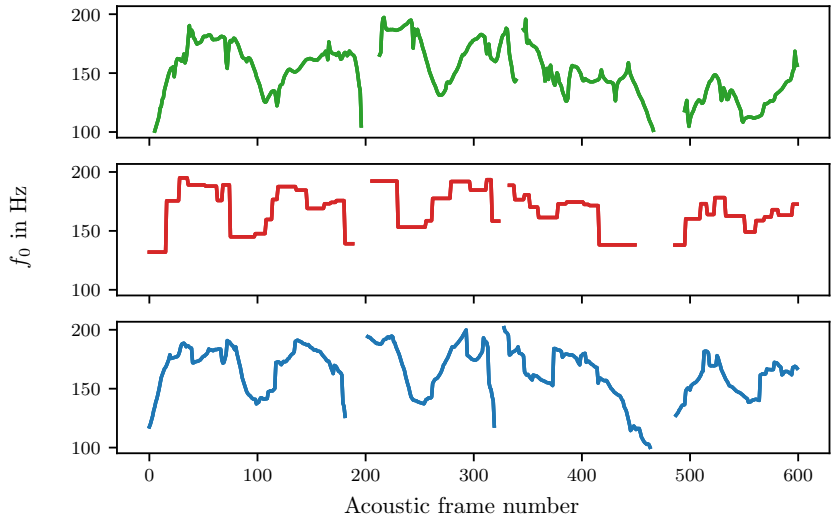}
    }
    \end{center}
    \captionsetup{justification=centering}
    \caption{Comparison of the Original $f_0$ contour (top), predicted $f_0$ contours before (middle) and after (bottom) frame position information embedding.}
    \label{comp_fo}
\end{figure}
The $f_0$ is modeled explicitly in the Prosody-TTS system to have control over it during the synthesis. $f_0$ estimator is a subtask that predicts the $f_0$ of each frame from the encoded representations, which are initially at linguistic phoneme rate. So, they are upsampled to acoustic frame rate by replicating them according to the corresponding phoneme duration and fed to the $f_0$ estimator. As the upsampled representations are identical for all the frames of a phoneme, the predicted $f_0$ is also identical for all the frames of the corresponding phoneme. As a result of this, $f_0$ contour looks piece-wise continuous, as shown in the figure \ref{comp_fo}. The resulting speech sounds unnatural as it uses constant $f_0$ for all the frames of a phoneme.

\begin{figure}[t]
 \centering
    \includegraphics{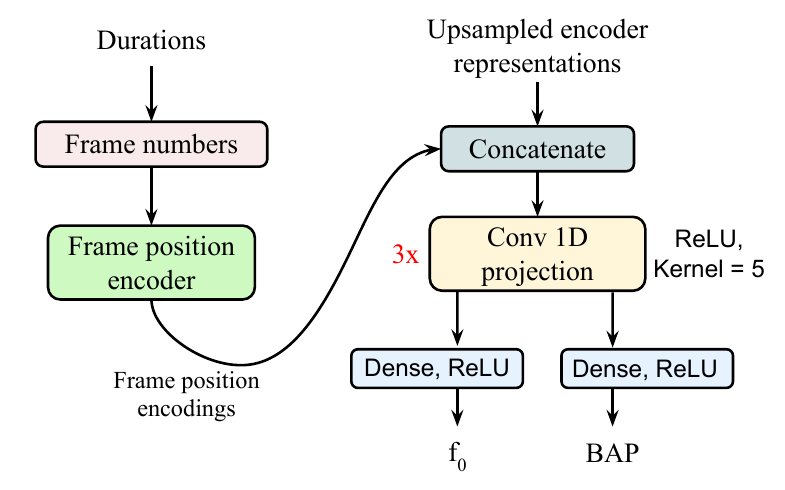}
 \caption{Architecture of the $f_0$ estimator module}
 \label{f0_est_arch}
\end{figure}

In order to generate a continuous $f_0$ contour, we need to get the unique representations for each frame of a phoneme. So, frame position information is concatenated to the upsampled encoder representations. First, we obtain the total number of frames of each phoneme by dividing its duration by frameshift. The position of the frame from both the beginning and the end of a phoneme is encoded with a 16-dimensional non-learnable vector. This 32-dimensional vector, named frame position information, is concatenated to the upsampled encoder representations to make them unique. These concatenated representations are fed to the convolutional projection layers to predict the $f_0$ and BAP. BAP is defined as the ratio between the periodic energy and the total energy in the given frequency band, where periodic energy is obtained by taking the sum of energies of all the harmonic components in the corresponding band. The correlation between the $f_0$ and BAP helps in using a single module for predicting both $f_0$ and BAP. The predicted $f_0$ contour looks continuous, as shown in the bottom plot of the figure \ref{comp_fo} when the frame position information is incorporated.

The $f_0$ estimator consists of three one-dimensional convolutional layers with a kernel size of 5, as shown in the figure \ref{f0_est_arch}. So, the effective receptive field is 13 frames, which indicates that 13-frame context information is used to model the $f_0$ prediction. The logarithm of $f_0$ is modeled since it follows the Gaussian distribution \cite{Merlin}, and thus we can use mean square error (MSE) as the loss function. The default $f_0$ value of the silence and unvoiced frames is set to 1Hz to ensure that the $\log f_0$ is non-negative. We have used the ReLU activation function as $\log f_0$ and BAP are positive. The loss functions related to $f_0$ and BAP are given by
\begin{equation}
    \mathcal{L}_{f_0} =  \frac{1}{K} \sum_{k=1}^K  |\log {f_0}_k - \log \widehat{f_0}_k|^2
\end{equation}
\begin{equation}
    \mathcal{L}_{BAP} =  \frac{1}{K} \sum_{k=1}^K  | bap_k -  \widehat{bap}_k|^2
\end{equation}
where $f_0$ and $\widehat{f_0}$ are the original and predicted $f_0$, respectively, K is the total number of frames in the utterance, bap and $\widehat{bap}$ are the original and predicted BAP, respectively.

\subsection{Acoustic decoder module}

Most state-of-the-art speech synthesis models use an autoregressive-based decoder, where the prediction is sequential. The autoregressive models require significantly more inference time than the non-autoregressive models, which are at least as accurate as the autoregressive models \cite{ar_vs_nonar}. The accuracy of speech synthesis models can be improved by incorporating the knowledge of human speech production mechanism. So, we choose to use a non-autoregressive acoustic decoder module and to overcome the limitation of accuracy, we incorporate the knowledge of speech production mechanism. The module takes the predicted $f_0$ and the bottleneck representations from the $f_0$ estimator as the input. Instead of supplying the $f_0$ input as a single neuron, we use a 32-dimensional continuous vector encoding for each $f_0$ value to match the magnitude range of the $f_0$ input with the other inputs \cite{high_fo}. We can also use the continuous wavelet transform (CWT) to convert the $f_0$ contour to higher dimension representation \cite{CWT}. The predicted $f_0$ is rounded to the nearest integer, and its corresponding 32-dimensional encoding is obtained from the stored lookup table. We must ensure that the range of the lookup table should also include the minimum and maximum values of the modified $f_0$ like x0.75, x1.5, etc. The lookup table may be replaced with the sinusoidal positional encoding, as the nearest $f_0$ values will have closer encodings. The extracted $f_0$ encodings are then used for conditioning the mel spectrogram prediction.

\begin{figure}[t]
 \captionsetup{justification=centering}
 \centering
    \includegraphics{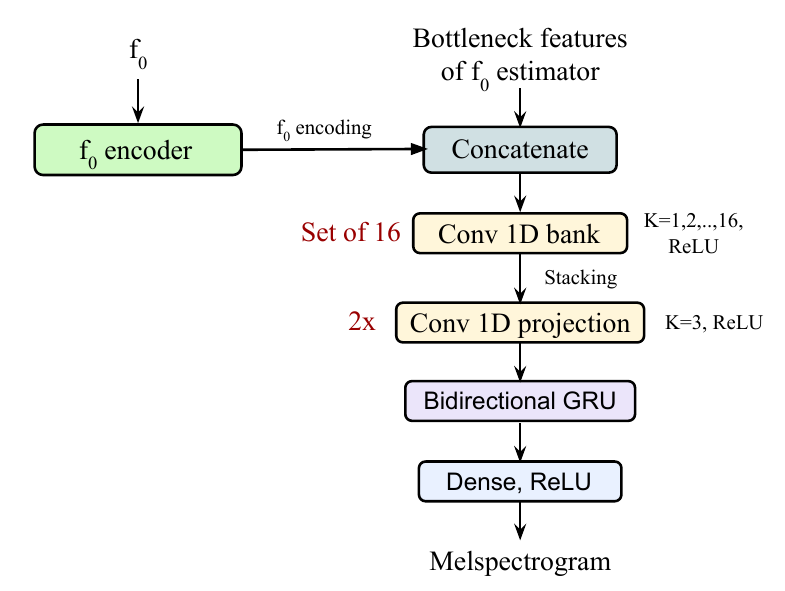}
 \caption{Architecture of the acoustic decoder module. K denotes the kernel size}
 \label{acst_dec_arch}
\end{figure}

The architecture of the acoustic decoder module is shown in the figure \ref{acst_dec_arch}. It consists of a bank of one-dimensional convolution layers, followed by a bidirectional GRU to estimate the mel spectrogram. We use a set of 16 convolutional filters in the bank, where the $k^{th}$ filter in the bank has the kernel size of k. The output of each convolution filter in the bank is stacked horizontally, projected into the lower dimension using the convolution projection layer, and fed to the bidirectional GRU. The effective receptive field obtained with the convolutional layers is 21 frames, with ten frames on either side of the current frame. The formant trajectories in the magnitude spectrum are captured through this receptive field, and ReLU activations are used to ensure that the predicted mel spectrogram is positive. The task of mel spectrogram prediction uses MSE as the loss function in the logarithmic domain, which is given by:
\begin{equation}
    \mathcal{L}_{mel} = \frac{1}{K} \sum_{k=1}^K |\log Y_k - \log \widehat{Y}_k|^2
\end{equation}
where $Y_k$ and $\widehat{Y}_k$ are the original and predicted mel spectrograms of the $k^{th}$ frame, respectively, K is the total number of frames in the utterance. The total loss of the auxiliary tasks associated with the acoustic parameters is given by
\begin{equation}
    \mathcal{L}_{acst} = \mathcal{L}_{f_0} + \mathcal{L}_{BAP} + \mathcal{L}_{mel}
\end{equation}
The predicted mel spectrogram from the acoustic decoder module is fed to the neural vocoder along with the excitation signal, which is generated from the $f_0$ and BAP.


\subsection{Neural Vocoder}
\label{sec:vocoder}
\begin{figure}[t]
 \centering
    \includegraphics{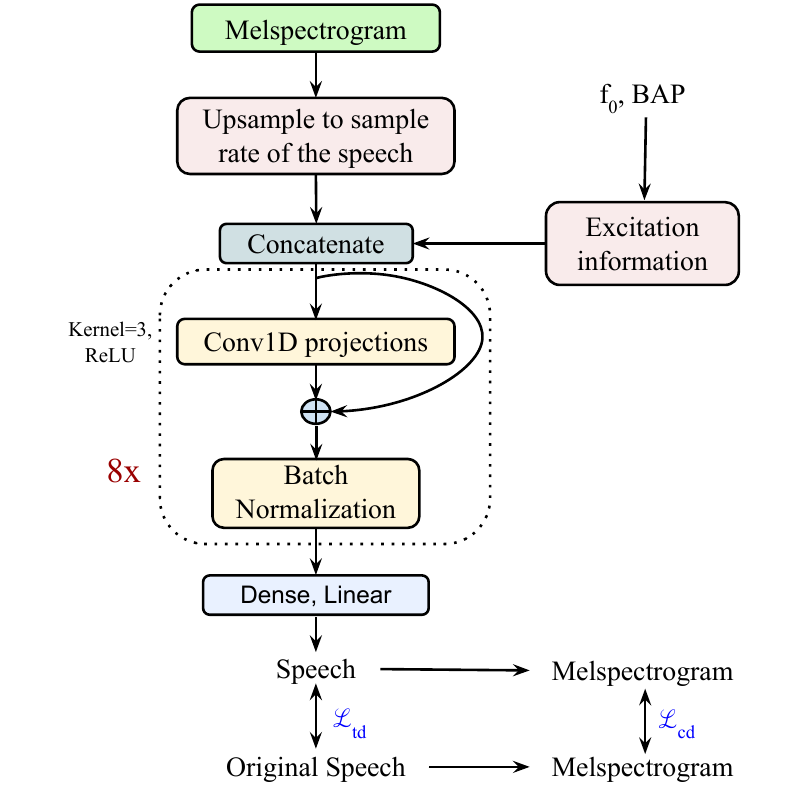}
 \caption{Architecture of the neural vocoder}
 \label{voc_arch}
\end{figure}
\begin{figure}[t]
    \captionsetup{justification=centering}
    \resizebox{\textwidth}{!}{%
    \centering
        \input{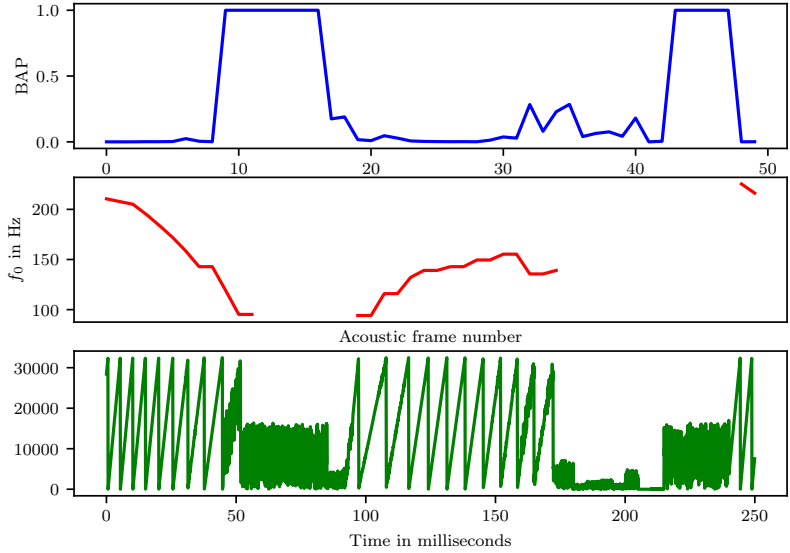}
    }
    \caption{Band Aperiodicity (BAP) curve (top), fundamental frequency ($f_0$) curve (middle) and the generated excitation signal from $f_0$ and BAP}
    \label{exc}
\end{figure}

The architecture of the neural vocoder is shown in the figure \ref{voc_arch}. We use a slightly modified version of Waffler \cite{waffler}, a non-autoregressive neural vocoder, to generate the speech signal from the mel spectrogram and excitation input. An excitation signal is generated using the $f_0$ and BAP, obtained from the $f_0$ estimator. First, the sequence of fundamental periods (1/$f_0$) is converted into the sample domain by multiplying with the sample rate (sample rate/$f_0$ sequence). A sawtooth pulse is created for each value in the sequence, and concatenation of them will result in a sawtooth pulse train. Gaussian noise, with BAP value as the standard deviation, is generated at 16kHz sample rate and added to the sawtooth pulse train to obtain the excitation signal, shown in the figure \ref{exc}. The predicted mel spectrogram, which is at the acoustic frame rate, is upsampled to the sample rate of the speech signal and fed to the convolutional neural network along with the excitation signal to generate the speech signal. Similar to Waffler \cite{waffler}, the neural vocoder is trained using the combination of time-domain loss ($\mathcal{L}_{td}$) and cepstral-domain loss ($\mathcal{L}_{cd}$), which are given by

\begin{equation}
    \mathcal{L}_{td} = \frac{1}{T} \sum_{n=1}^T |s(n) - \widehat{s(n)}|^2
\end{equation}

\begin{equation}
    \mathcal{L}_{cd} = \frac{1}{K} \sum_{k=1}^K |\log S_k - \log \hat{S_k}|^2
\end{equation}
where $s(n)$ is the original speech sample, $\widehat{s(n)}$ is the predicted speech sample from the vocoder, T is the total number of samples in the speech signal, K is the total number of frames, $S_k$ and $\hat{S_k}$ are the original and predicted mel spectrograms of $k^{th}$ frame extracted from the $s(n)$ and $\widehat{s(n)}$, respectively. The loss function of the vocoder is considered as the combination of $\mathcal{L}_{td}$ and $\mathcal{L}_{cd}$.

\begin{equation}
    \mathcal{L}_{vocoder} = \lambda \mathcal{L}_{td} + (1 - \lambda) \mathcal{L}_{cd}
\end{equation}
where $\lambda$ is the hyper-parameter, which decides the importance given to each loss. The model is trained in an end-to-end fashion using auxiliary learning with the primary task of predicting speech from the text input. The total loss of the model is the sum of loss functions of all the subtasks.
\begin{equation}
    \label{total_loss}
    \mathcal{L}_{total} = \mathcal{L}_{dur} + \mathcal{L}_{acst} + \mathcal{L}_{vocoder}
\end{equation}

\section{Speech Synthesis Experiments}
\label{sec:Exper_setup}
The Prosody-TTS system is trained on the Telugu language speech data from the IndicTTS database \cite{dataset}, collected from a male speaker. The database consists of text and the corresponding speech file at a 16kHz sample rate. It has around 2450 utterances, which corresponds to 4 hours of speech data. There are a total of 58 unique phonemes along with the punctuation marks in the database. Twenty utterances containing 28 percent of unseen words are held out to evaluate the performance of the trained model. We can also extend the proposed method to other languages since we do not require a language-specific text analyzer to extract the linguistic context information. Instead, the context information and the coarticulation constraints are learned from the data.

\subsection{Feature extraction}
\label{sec:feat_ext}
The text and its corresponding speech signal are required to train the Prosody-TTS system. As the model is trained using supervised auxiliary learning, labeled data like duration, $f_0$, BAP, and mel spectrogram are required. The ground truth durations of the phonemes are obtained using the state-level forced alignment \cite{htk_book}. Then the total duration of the phoneme is obtained by taking the sum of all the state durations of the corresponding phoneme. The acoustic parameters $f_0$ and BAP are obtained by using the WORLD vocoder \cite{World}. 80-dimensional mel spectrograms are extracted from speech signal with 2048 point discrete Fourier transform (DFT).  All the acoustic parameters are extracted with a frame length of 25ms and a frameshift of 5ms.

\subsection{Model Details and Training}
\label{sec:model_det}
\begin{figure}[t]
    \resizebox{\textwidth}{!}{%
    \centering
        \input{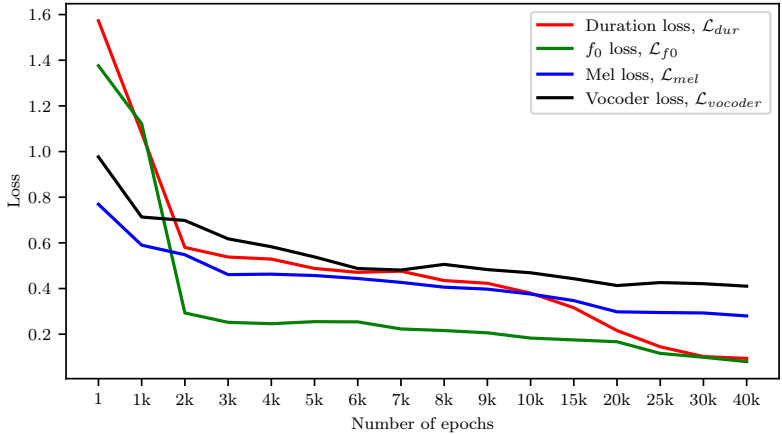}
    }
    \captionsetup{justification=centering}
    \caption{Training loss curves of different auxiliary loss functions VS number of epochs.}
    \label{loss_curves}
\end{figure}

The parameters of the Prosody-TTS model are shown in the table \ref{nw_para}. The model is trained with a batch size of 16. As shown in the equation (\ref{total_loss}), the total loss of the model is the combination of loss functions of individual subtasks, which is minimized using the Adam optimizer \cite{Adam} with $\beta_1 = 0.9$ and $\beta_2 = 0.999$. We use an initial learning rate of 0.002, which is reduced after 50000 iterations to a learning rate of $10^{-4}$. The convergence of the loss functions of all the subtasks is shown in the figure \ref{loss_curves}. The implementation\footnote{https://github.com/siplabiith/Prosody-TTS  \newline
  \hspace*{1.5em}  Currently, the repository is password protected. Use the password \textbf{tts@SipLab}. It will be made public once the paper is published.} of the model is made available online to encourage reproducibility.

\begin{table}[htbp]
  \caption{Network parameters of the proposed model, K denotes kernel}
  \label{nw_para}
  \centering
  {%
   \begin{tabular}{c|c|c}
    \hline
   \textbf{Module} & \textbf{Layer}  & \textbf{Parameters} \\
    \hline
   \multirow{5}{*}{\parbox{4cm}{\textbf{Text Encoder \& \\ Duration estimator}}} 
   &  Character embedding & 128D \\
   &  Conv1D bank (8)&  128, K=1,2,..,8, ReLU \\
   & 2 X Conv1D projections    & 128, K=3, ReLU \\
   & Bidirectional GRU    &  128 \\
   & Output(Duration)    &  1 \\
    \hline
  \multirow{3}{*}{\textbf{$f_0$ estimator}} 
   & Frame positional encodings & 16D, 16D \\
   & 3 X Conv1D projection    & 128, K=5, ReLU \\
   & $f_0$, BAP    &  1, 1 \\
   \hline
  \multirow{5}{*}{\textbf{Acoustic decoder}} 
   & $f_0$ encodings    & 32D \\
   &  Conv1D bank (16) &  128, K=1,2,..16, ReLU \\
   & 3 X Conv1D projection    & 128, K=3, ReLU \\
   & Bidirectional GRU    &  128  \\
   & Mel spectrogram    &  80 \\
   \hline
  \multirow{6}{*}{\textbf{Neural vocoder}} 
   & Excitation input & 16000  \\
   & Mel spectrogram input &  80  \\
   & Upsampling layer &  x200 \\
   & 8 X Conv1D projection    & 64, K=3, ReLU \\
   & Output (Speech samples)  & 16000  \\
   \hline
  \end{tabular}
  }
\end{table}

\subsection{Results}
\label{sec:Results}
We first evaluate the amount of context information required to model the phoneme duration by experimenting with different convolution filters like 2, 4, 8, and 16 in the convolution bank of the text encoder module. The context information depends on the receptive field of the text encoder module, which depends on the number of convolutional filters. In all the experiments, the convolution bank is followed by two projection layers with a kernel size of 3. The performance is evaluated by predicting the phoneme durations and estimating the Root mean square error (RMSE), Mean absolute error (MAE), Pearson's correlation coefficient (PCC), and the results are tabulated in the table \ref{cont_info}. We can notice that the performance of the model is degraded when the receptive field is very low, as the context information used to predict the duration of the phoneme is lacking in such cases. The performance of the model also degrades when we use a high receptive field like 20, as there are fewer examples in the training data with such higher context dependency, and the model gets over-fitted. So, we found that the model performs better for the moderate receptive field of 12, which is obtained when eight convolutional filters are used in the bank.

\begin{table}[htbp]
  \caption{Evaluation of the context information required for duration modeling}
  \label{cont_info}
  \centering
  \begin{tabular}{c|c|c|c|c}
    \hline
  \thead{\textbf{No. of filters in} \\ \textbf{Convolution bank} }  & \thead{\textbf{Receptive} \\ \textbf{field}}  & \textbf{RMSE}  & \textbf{MAE} & \textbf{PCC}\\
    \hline
    2  &  6   &  4.653   & 2.787  & 0.802 \\
    \hline
    4  &  8   &  4.609  & 2.775  & 0.809  \\
    \hline
    \textbf{8}  &  \textbf{12}   & \textbf{4.437}   & \textbf{2.694}  & \textbf{0.818}  \\
    \hline
    16 &  20   & 4.813   & 2.996  & 0.787  \\
    \hline
  \end{tabular}
\end{table}

\subsubsection{Subjective Evaluation}

\begin{figure}[t]
    \resizebox{\textwidth}{!}{%
    \centering
        \input{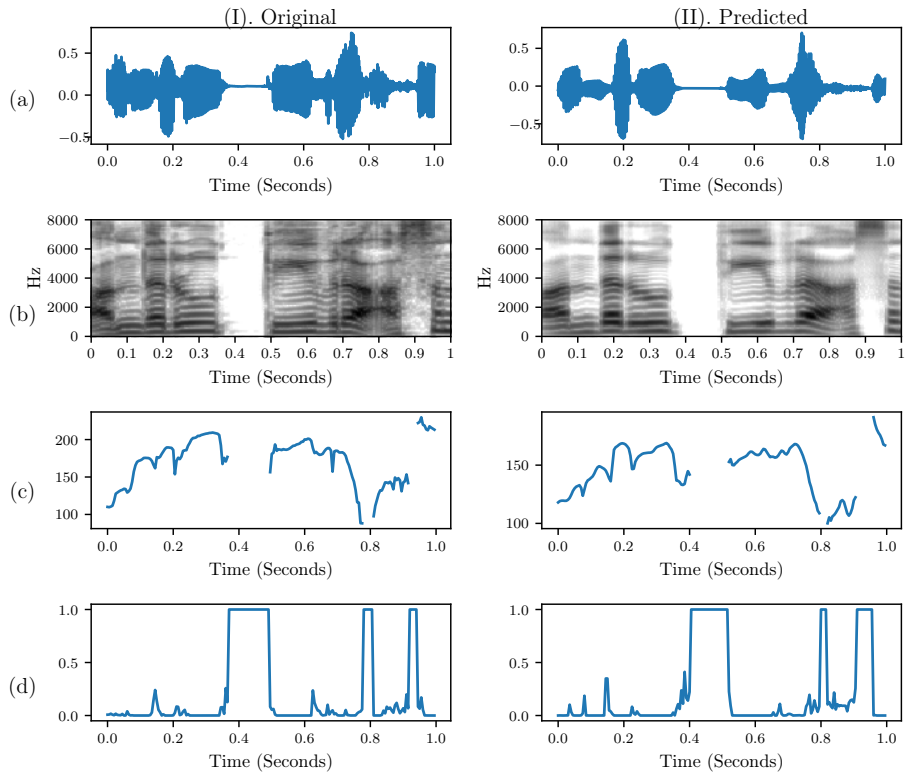}
    }
    \captionsetup{justification=centering}
    \caption{Comparison between the original and predicted speech along with the acoustic features. (a): Speech (b): mel spectrogram, (c): $f_0$ in Hz and (d): BAP }
    \label{comp_4}
\end{figure}

The comparison between the original and predicted speech along with its acoustic features is shown in the figure \ref{comp_4}. The performance of the Prosody-TTS model is evaluated using the 20 sentences, which are held out for testing. Text from this test data is converted into the character embeddings and fed to the trained Prosody-TTS model to synthesize the corresponding speech signal. The quality of the synthesized speech\footnote{Synthesized samples are available at: https://siplabiith.github.io/prosody-tts.html} is rated using the MOS by 30 native Telugu language speakers. We compare the Prosody-TTS model with different parametric methods and end-to-end models such as Parametric+WORLD, Parametric+WaveNet, WaveNet \cite{Wavenet}, and Tacotron \cite{Tacotron}. For the Parametric+WORLD system, we trained the Merlin speech synthesis tool \cite{Merlin} and WORLD \cite{World} is used as the vocoder. The WORLD vocoder is replaced with the WaveNet vocoder in Parametric+WaveNet, which is trained individually. Similarly, end-to-end speech synthesis models like WaveNet and Tacotron are also trained with the same Telugu language male speaker data. We can observe from the table \ref{mos_comp} that the proposed model achieves better MOS even when compared to the autoregressive models like WaveNet and Tacotron, with very low inference time. The real time factor (RTF) is used as the evaluation parameter for the inference time, which is defined as the ratio of time taken to synthesize the speech and the duration of the speech synthesized. The RTFs for all the models are evaluated on the Nvidia GeForce RTX 2080 Ti GPU.

\begin{table}[htbp]
  \captionsetup{justification=centering}
  \caption{The MOS of different models with 95\% confidence interval and the corresponding Real time factors}
  \label{mos_comp}
  \centering
   \begin{tabular}{c|c|c}
    \hline
   \textbf{Model} & \textbf{MOS} & \textbf{RTF}  \\
    \hline
   Parametric + WORLD  &  2.76 $\pm$ 0.13  & \textbf{0.24} \\ 
   Parametric + WaveNet  &   3.16 $\pm$ 0.11 & 232.1  \\ 
   WaveNet  &  3.46 $\pm$ 0.08   & 254.6 \\ 
   Tacotron  &  3.87 $\pm$ 0.07 & 183.2 \\ 
   Prosody-TTS  &  \textbf{4.06 $\pm$ 0.08}  & 0.37 \\ 
    \hline
  \end{tabular}
\end{table}

\subsubsection{Objective Evaluation}
We have conducted the quantitative analysis to evaluate the performance of the Prosody-TTS using the predicted $f_0$ and durations. Gross pitch error (GPE) \cite{gpe_vde}, Voicing decision error (VDE) \cite{gpe_vde}, and $f_0$ frame error (FFE) \cite{ffe} are used to compare the performance of different systems based on $f_0$. GPE is defined as:
\begin{equation}
    GPE = \frac{N_v^{f0err}}{N_v} \times 100
\end{equation}
where $N_v$ is the total number of voiced frames in the utterance and $N_v^{f0err}$ is the number of voiced frames in which the predicted $f_0$ doesn't lie in the range of $\pm20\%$ of the original $f_0$ value. VDE is defined as:
\begin{equation}
    VDE = \frac{N_v^{err} + N_u^{err}}{N} \times 100
\end{equation}
where N is the total number of frames in the given utterance, $N_v^{err}$ is the number of voiced frames predicted as unvoiced, and $N_u^{err}$ is the number of unvoiced frames predicted as voiced. WORLD vocoder is used to extract the $f_0$ and voiced/unvoiced labels from the synthesized signals of all the systems. $f_0$ frame error uses three different types of errors, which is defined as:
\begin{equation}
    \begin{split}
        FFE & = \frac{\text{No. of error frames}}{N} \times 100 \\
        & = \frac{N_v^{f0err} + N_v^{err} + N_u^{err}}{N} \times 100 \\
        & = \frac{N_v}{N} \times GPE + VDE
    \end{split}
\end{equation}
Thus FFE is a combination of GPE and VDE. Similarly, duration related errors like RMSE, MAE, and PCC are evaluated to analyze the predicted durations. The obtained quantitative results of the different systems are compared in the table \ref{erros_comp}.

\begin{table}[htbp]
  \caption{Comparison of the different errors related to $f_0$ and duration.}
  \label{erros_comp}
  \centering
   \begin{tabular}{c|c|c|c|c|c|c}
    \hline
    \multirow{2}{*}{\textbf{Model}} & \multicolumn{3}{c|}{\textbf{$f_0$ errors (\%)}}  & \multicolumn{3}{c}{\textbf{Duration errors}}\\ 
    \cline{2-7}
   & \textbf{GPE} & \textbf{VDE} & \textbf{FFE} & \textbf{RMSE} & \textbf{MAE} & \textbf{PCC}\\
    \hline
    Parametric  & 22.13 & 7.16 & 23.29 & 7.104 & 5.039 & 0.728 \\ 
   Tacotron & 9.95 & 5.76 & 12.65 & - & - & - \\
   Prosody-TTS  & \textbf{3.72} & \textbf{2.96} & \textbf{4.23} & \textbf{4.381} &  \textbf{2.694} & \textbf{0.816} \\ 
    \hline
  \end{tabular}
\end{table}

\section{Prosody control}
\label{pros_cont}

Since the proposed system explicitly estimates the phoneme duration and f0, we can control them before estimating the acoustic parameters for speech synthesis. This is an effective way of controlling the prosody, as we operate at the feature level instead of operating at the waveform level. We can either modify the prosody or incorporate the desired/target prosody in the synthesized speech. Based on the type of control, they are classified as Prosody modification and Prosody incorporation. Prosody modification aims to alter the prosodic parameters without a target or reference prosody, for example, a constant factor multiplication to the duration and $f_0$. Prosody incorporation aims at incorporating the desired prosody in the synthesized signal. In this paper, we have evaluated the Prosody-TTS system on both the prosody modification and prosody incorporation tasks.

\subsection{Prosody modification}
In the Prosody modification task, the phoneme duration and $f_0$ are multiplied with a constant factor ranging from 0.5 to 1.5 at the feature level to modify the prosody. Although the prosodic parameters are multiplied with a constant factor, it has several applications like emotion conversion, expressive speech synthesis, etc. Govind et al. \cite{emo_conv_new} proposed an emotion conversion method based on modifying $f_0$ and duration of the neutral speech according to the multiplication factors obtained from the emotional speech data. The prosody modified speech signals are also evaluated using the MOS and compared with the neural parametric speech systems, as they have the provision for altering the prosody, which is shown in table \ref{mos_comp_prosody_mod}. Although the speech rate and the pitch are modified, there is no considerable degradation in the naturalness and intelligibility, which can be inferred from the obtained MOS. There is no substantial improvement in the MOS as the prosodic parameters are multiplied with a constant factor for the entire utterance. The proposed system can also be used for emotion conversion \cite{emo_conv_new} by obtaining the modification factors from the emotional speech data and using them for multiplying the prosodic parameters.

\begin{table}[htbp]
  \caption{The MOS of prosody modified speech with 95\% confidence interval}
  \label{mos_comp_prosody_mod}
  \centering
   \begin{tabular}{c|c}
    \hline
   \textbf{Model} & \textbf{MOS}  \\
    \hline
   Parametric + WORLD  &  2.61 $\pm$ 0.12   \\ 
   Parametric + WaveNet  &  3.14 $\pm$ 0.07  \\ 
   Prosody-TTS  &   \textbf{4.03 $\pm$ 0.06} \\ 
    \hline
  \end{tabular}
\end{table}

\subsection{Prosody incorporation}
In the prosody incorporation task, the reference prosodic parameters (duration and $f_0$) are extracted from the target/reference speech using forced alignment and WORLD vocoder, respectively. The text corresponding to the reference prosody is provided as input to the model. During the synthesis, the reference durations and $f_0$ are incorporated into the synthesized speech signal. We have evaluated the prosody incorporated speech using MOS, which are shown in the table \ref{mos_comp_prosody_incor}. We can observe the improvement in the MOS, as the reference prosody is incorporated during the synthesis. The prosody incorporation task has several applications like automatic dubbing (subtitle synthesis), speech-to-speech translation, voice conversion, etc.

\begin{table}[htbp]
  \captionsetup{justification=centering}
  \caption{The MOS of reference prosody incorporated speech with 95\% confidence interval}
  \label{mos_comp_prosody_incor}
  \centering
   \begin{tabular}{c|c}
    \hline
   \textbf{Model} & \textbf{MOS}  \\
    \hline
   Parametric + WORLD  &   2.94 $\pm$ 0.07  \\ 
   Parametric + WaveNet  &  3.32 $\pm$ 0.10  \\ 
   Prosody-TTS  &  \textbf{4.09 $\pm$ 0.09}  \\ 
    \hline
  \end{tabular}
\end{table}

As future work, we are developing an automatic dubbing (subtitle synthesis) system, where the subtitle text of a source video is synthesized in the target language and mixed with the source video. In this case, the duration of the synthesized signal should match with the source video. So, we should control the phoneme durations to align the synthesized signal with the events in the source video and to match the total duration. Subtitle synthesis has numerous applications in the education and entertainment sector.

\section{Conclusion and Future Work}
\label{sec:Conclusion}
In this paper, we proposed Prosody-TTS, an end-to-end speech synthesis system based on supervised auxiliary learning, which can control the prosody of the synthesized speech. The model combines the advantages of parametric and end-to-end speech synthesis systems. It explicitly models the phoneme duration and $f_0$ to have control over them during the synthesis. The system achieves a MOS of 4.08 with the naturalness close to the original speech signals and with a very low inference time as the model is non-autoregressive. The system can be incorporated in several applications like emotion conversion, automatic dubbing, and speech-to-speech translation tasks, where we should have control over the prosodic parameters (duration and $f_0$). Since the energy of the signal also contributes to the prosody, as a future work, we will find a way to incorporate the energy by conditioning on the desired emotion. We are also developing a subtitle synthesis system based on the Prosody-TTS, where the speech signal is synthesized in the target language from the subtitles of the source video with the desired source prosody.

\bibliographystyle{spmpsci}      
\bibliography{refs}   

%
%

\end{document}